\documentclass[aps,prl,twocolumn,groupedaddress,preprintnumbers,showpacs,showkeys]{revtex4}

\usepackage{amsmath}
\usepackage{scalefnt}
\usepackage{graphicx}
\graphicspath{{./figures/}}
\begin{document}
\title{Charm and Bottom Quark Masses: an Update} 
\author{K. G. Chetyrkin} 
\author{J. H. K\"uhn} 
\author{A. Maier} 
\author{P. Maierh\"ofer}
\author{P. Marquard}
\author{M. Steinhauser}
\affiliation{Institut f\"ur Theoretische Teilchenphysik,
Universit\"at Karlsruhe, Karlsruhe~Institute~of~Technology~(KIT), 76128
Karlsruhe, Germany} 
\author{C. Sturm}
\affiliation{Physics Department, Brookhaven National Laboratory,
Upton, New York 11973, USA}
\begin{abstract}
Using new four-loop results for the heavy quark vacuum 
polarization and new data for bottom quark production 
in electron-positron annihilation, an update on the 
determination of charm- and bottom-quark masses through 
sum rules has been performed. The previous result for 
the charm-quark mass, $m_c(3~{\rm GeV})= 0.986(13)~{\rm GeV}$, 
based on the lowest moment, is supported by the new 
results from higher moments which lead to consistent values 
with comparable errors. The new value for the bottom quark,
$m_b(10~{\rm GeV})= 3.610(16)~{\rm GeV}$, corresponding to
$m_b(m_b)= 4.163(16)~{\rm GeV}$, makes use both of the new 
data and the new perturbative results and is consistent 
with the earlier determination. 
\end{abstract}
\pacs{11.55.Hx 12.38.Bx 14.65.Dw 14.65.Fy}
\keywords{precise charm and bottom quark mass, sum rules, perturbative calculations}

\preprint{TTP09-16, SFB/CPP-09-49}

\maketitle

\section{I. Introduction}
\label{sec:intro}

The precise determination of charm and bottom quark masses has always been an 
important task both for theory and experiment. The most precise values have
been obtained \cite{Kuhn:2007vp} from an analysis of the ITEP sum rules
\cite{Novikov:1977dq} 
(for reviews see Refs.~\cite{Reinders:1985sr,Colangelo:2000dp,Ioffe:2005ym}), 
combining data for the heavy quark production cross section in
electron-positron collision with dispersion relations and a four-loop
evaluation of the vacuum polarization induced by the heavy quark
current. In this letter,
we present an update of these results. We will include data recently
published by the BABAR collaboration \cite{:2008hx} and make use of new perturbative 
results which replace the estimates for the four-loop coefficients of 
higher moments used in the earlier publications.

\section{II. Analytic results}
\label{sec:res}

Our determination of the heavy quark masses follows closely
Refs.~\cite{Kuhn:2007vp,Kuhn:2001dm,Kuhn:2002zr}. 
It is based on the direct comparison of the
theoretical and experimental evaluations of the contributions to the
derivatives of the polarization function $\Pi_Q(q^2)$, the former evaluated in perturbative QCD,
the latter through moments of the measured cross section for heavy quark
production in electron-positron annihilation. Using dispersion relations, 
the moments of $R_Q$~\footnote{For the precise definition of $R_Q$, in
  particular the treatment of gluon 
  splitting into $Q\bar Q$, the subtraction of singlet contributions, and the
  role of nonperturbative terms in the case of charm quarks we refer to
  Ref.~\cite{Kuhn:2007vp}.} 
\begin{equation}
  {\cal M}_n \equiv \int \frac{{\rm d}s}{s^{n+1}} R_Q(s)
  \,,
  \label{eq:Mexp}
\end{equation}
can be related to the derivatives of the vacuum polarization function at $q^2=0$
\begin{equation}
  {\cal M}_n = \frac{12\pi^2}{n!}
  \left(\frac{{\rm d}}{{\rm d}q^2}\right)^n
  \Pi_Q(q^2)\Bigg|_{q^2=0}
  \,.
\label{eq:Mtheo}
\end{equation}
 In its domain of
analyticity $\Pi_Q(q^2)$ can be cast into the form
\begin{equation}
  \Pi_Q(q^2) = Q_Q^2 \frac{3}{16\pi^2} \sum_{n\ge0}
                       \bar{C}_n z^n
  \,,
  \label{eq:pimom}
\end{equation}
with $z=q^2/(4m_Q^2)$. Here  $m_Q=m_Q(\mu)$ is the heavy quark
mass with charge $Q_Q$ in the $\overline{\rm{MS}}$ scheme at the scale
$\mu$. The coefficients 
$\bar C_n$ depend on $\alpha_s$ and 
on the heavy quark mass through logarithms of the form $l_{m_Q} = \ln
(m_Q^2(\mu)/\mu^2)$. Equating theoretically calculated and
experimentally measured moments, the heavy quark mass is given by 
\begin{equation}
  m_Q(\mu) = \frac{1}{2}
  \left( \frac{9 Q_Q^2\bar{C}_n}{4 {\cal M}_n^{\rm exp}}\right)^{1/(2n)} 
  \,.
  \label{eq:m_Q}
\end{equation}
As a perturbative series the coefficients $\bar{C}_n$ can be written as
\begin{equation}
  \begin{split}
    \bar{C}_n =&\, \bar{C}_n^{(0)} + \frac{\alpha_s(\mu)}{\pi} \left(
      \bar{C}_n^{(10)} + \bar{C}_n^{(11)}l_{m_Q} \right)
    \\&
    +\left(\frac{\alpha_s(\mu)}{\pi}\right)^2 \left( \bar{C}_n^{(20)}
      +\bar{C}_n^{(21)}l_{m_Q}
      + \bar{C}_n^{(22)}l_{m_Q}^2 \right)
    \\&
    + \left(\frac{\alpha_s(\mu)}{\pi}\right)^3 \left( \bar{C}_n^{(30)} 
      + \bar{C}_n^{(31)}l_{m_Q} 
      + \bar{C}_n^{(32)}l_{m_Q}^2 
    \right.\\&\left.\mbox{}\qquad
      + \bar{C}_n^{(33)}l_{m_Q}^3 \right)
    + \ldots \,.
  \end{split}
  \label{eq:cn}
\end{equation}
The terms of order $\alpha_s^2$ were evaluated up to $n=8$ in 
Refs.~\cite{Chetyrkin:1995ii,Chetyrkin:1996cf,Chetyrkin:1997mb} (and
recently in Refs.~\cite{Boughezal:2006uu,Maier:2007yn} even up to n=30). The four loop
contributions 
to $\bar C_0$ and $\bar C_1$ were calculated in
Refs.~\cite{Chetyrkin:2006xg,Boughezal:2006px}.
For the higher moments the analysis of \cite{Kuhn:2007vp} was based on estimates 
for $\bar C_n^{(30)}$ with $n=2,~3,~4$, which lead to an additional
uncertainty in the 
mass determination. Recently 
the exact results for the second~\cite{Maier:2008he} and
third~\cite{Maier:2009} moments were  
obtained. Combining these coefficients with additional information on 
the threshold and the high-energy behaviour and using the analyticity of
$\Pi_Q(q^2)$ and Pad\'e approximations, fairly precise numerical results 
were obtained~\cite{Kiyo:2009} for the higher coefficients up to $n=10$. (For
an earlier analysis
along similar lines see Ref.~\cite{Hoang:2008qy}.)   
For the lowest four moments the four loop coefficients $\bar C_n^{(30)}$ are
listed in Tab.~\ref{tab:1} both for the charm and the bottom quark. All other
coefficients relevant for $n=1$ to $4$ can be found
in Tabs.~4 and~9 of \cite{Kuhn:2007vp}.
It should be emphasized that these results are well within the estimates used
in the analysis of \cite{Kuhn:2007vp}, also shown in Tab.~\ref{tab:1}. The
impact of these new results on the quark mass determination will be  
studied below. 

\begin{table}
  \centering
{\scalefont{0.8}
  \begin{tabular}{c||c|c|c|c}
    $n$  &  1 & 2 & 3 & 4\\\hline \hline
    charm &$-5.6404$&$-3.4937$ &$-2.8395$ & $-3.349(11)$ \\\hline
    \begin{minipage}{2.5cm}
      \centering
      lower $||$ upper limits\\\vspace{1mm}
    \end{minipage}
& ---&  $- 6.0\ ||\ 7.0$ & $-6.0\ ||\ 5.2$ &
    $-6.0\ ||\ 3.1$\\\hline \hline
    bottom &$-7.7624$ &$-2.6438$ &$-1.1745$ &$-1.386(10)$ \\\hline
\begin{minipage}{2.5cm}
      \centering
      lower $||$ upper limits
    \end{minipage}
     & --- & $- 8.0\ ||\ 9.5$ & $-8.0\ ||\ 8.3$ &
    $-8.0\ ||\ 7.4$\\
  \end{tabular}
}
  \caption{New results for the coefficients $\bar{C}_n^{(30)}$ in
    comparison with previous upper and lower limits as used in
    Ref.~\cite{Kuhn:2007vp}. For less precise numerical results of
    $\bar{C}_n^{(30)}$ for $n=3$ and $n=4$ see Ref.~\cite{Hoang:2008qy}.}
  \label{tab:1}
\end{table}


\section{III. Bottom Production Close to Threshold}
\label{sec:bot_prod}

The determination of the bottom quark mass, as performed in
\cite{Kuhn:2007vp,Kuhn:2001dm} relies heavily on the precise measurement of 
 $R=\sigma(e^+e^-\to {\rm hadrons})/\sigma_{\text{pt}}$
 (with $\sigma_{\text{pt}} =  \frac{4\pi\alpha^2}{3s}$),
which enters the moments as defined above. 
Specifically, it is the contribution from the heavy quark current
denoted as $R_b$ with the light quark contribution
subtracted.
It is convenient to split the integration region 
into three pieces: The lowest region covering the narrow 
resonances, an intermediate ``threshold'' region between 10.62~GeV
and 11.24~GeV, and the perturbative region above 11.24~GeV, where the
measurement is replaced by the perturbative QCD prediction. The choice
of 11.24~GeV corresponds to the upper end of the energy range covered by a
CLEO measurement 
more than 20 years ago \cite{Besson:1984bd}. It also coincides approximately
with the 
energy reach of a recent BABAR measurement \cite{:2008hx}. In the analysis of \cite{Kuhn:2007vp},
$\Upsilon(4S)$ with its mass
$M_{\Upsilon(4S)}=10.5794(12)$~GeV and width $\Gamma_{\Upsilon(4S)}=20.5$~MeV
has been considered together 
with the three lower, narrow resonances and thus the continuum part of the
bottom-cross section was taken from 10.62 GeV upwards. Until recently the 
only measurement in the threshold region has been the one from the CLEO
collaboration, which quotes a systematic error of about 6\%. No radiative
corrections had been applied. In Ref.~\cite{Kuhn:2007vp} it has been argued, that a
normalization factor 1/1.28 is necessary to reconcile these data
with more recent and more precise CLEO results below the $\Upsilon(4S)$-resonance 
and with perturbative QCD at the high end. These ``rescaled'' data
were the basis of the subsequent extraction of the bottom quark mass.
However, in view of these uncertainties an overall systematic error of
10\% was attributed to the contribution of the moments from this region.
Thus, although this contribution to the moments is relatively small, its
impact on the error was larger or equal than the one from the other 
two regions combined.

Recently a measurement of $R_b$ in the energy region between 10.54~GeV and
11.20~GeV was performed by the BABAR collaboration
with significantly improved statistics and with a correlated systematic
error between 2.5\% and 3\% \cite{:2008hx}. In principle this should allow an independent
determination of the contribution to the moments with significantly
reduced systematic error.
However, no radiative corrections were applied to the published data and the
radiative tails of the four lower $\Upsilon$ resonances were included in the
quantity denoted $R_b$. In the following we describe the procedure used
to obtain the contribution to the moments from these data.

In a first step we subtract the radiative
tail of the $\Upsilon(1S)$, $\Upsilon(2S)$ and $\Upsilon(3S)$
resonances, which is explicitly given in Ref.~\cite{:2008hx}.
Subsequently we subtract the radiative tail of the $\Upsilon(4S)$ resonance. For the resonance
shape we use a Breit-Wigner function with an electronic width of $\Gamma_{ee}(\Upsilon(4S))=0.272$~keV 
and an energy-independent total width of
$\Gamma_{\text{tot}}(\Upsilon(4S))=20.5$~MeV \cite{PDG}. For the radiator
function  
$G(z)$ we take the functional dependence as used in \cite{Chetyrkin:1996tz}, based on 
the resummed NNLO result of \cite{Jadach:1988gb}:
\begin{equation}
G(z) = \beta (1-z)^{\beta-1}\,{\rm e}^{\delta_{\text{yfs}}}\,F\,\left(
 \delta_C^{V+S} + \delta_C^H \right)\,,
\label{eqGc}
\end{equation}
with
\begin{align}
\beta =& \frac{2\alpha}{\pi}(L-1)\,,\qquad L = \ln\frac{s}{m_e^2}\,,
\qquad F = \frac{{\rm e}^{-\beta\gamma_E}}{\Gamma(1+\beta)}\,,\nonumber\\
\delta_{\text{yfs}} =& \frac{\alpha}{\pi} \left( \frac{L}{2} - 1 + 2\zeta(2) 
 \right)\,,\\
\delta_c^{V+S} =& 1+\frac{\alpha}{\pi}(L-1)+\frac{1}{2}
 \left(\frac{\alpha}{\pi}\right)^2 L^2\,,\nonumber\\
\delta_C^H =& -\frac{1-z^2}{2}+\frac{\alpha}{\pi}L\left[-\frac{1}{4}
 \left(1+3z^2\right)\ln z -1+z\right]\,.\nonumber
\end{align}

The remainder $\hat{\sigma}$ corresponds to the continuum cross section
distorted by initial-state radiation (ISR) and modified by vacuum polarization. It
is related to $\sigma$, the cross section without ISR, through
\begin{equation}
\hat\sigma(s) = \int_{z_0}^1 {\rm d}z\,G(z)\,\sigma(sz)\,,
\label{eqisrint}
\end{equation}
where the lower bound of the integration is given by $z_0 = (10.62
\text{GeV})^2/s$ corresponding to the point where the continuum cross
section (after subtraction of the $\Upsilon(4s)$ resonance) vanishes.

Given $\hat \sigma$, we can solve for $\sigma$ in an iterative way as
follows: Let us define $\delta G(z) \equiv G(z) -\delta(1-z)$ and
evaluate a successive series of approximations,
\begin{equation}
  \label{eq:sigma_i}
  \sigma_i = \sigma_{0} -\int_{z_0}^{1}{\rm d}z 
  \delta G(z) \sigma_{i-1}(sz)\,,
\end{equation}
using as starting point $\sigma_0=\hat{\sigma}$.
The difference between $\sigma_i$ and $\sigma$ can be estimated by
evaluating Eq. (\ref{eqisrint}) with $\sigma_i$ in place of $\sigma$.
After five iterations the resulting function differs from $\hat{\sigma}$
by less than $0.5\%$.

Finally, the effect of the vacuum
polarization must be taken into account and the result is normalised
relative to the point cross section,
\begin{equation}
  \label{eq:R_b}
  R_b=\sigma \frac{3s}{4\pi\alpha^2(s)}\,.
\end{equation}

The integration
region in Eq.~\eqref{eqisrint} covers the energy interval between 10.62~GeV
and 11.24~GeV, whence a constant value $(\alpha/\alpha(s))^2=0.93$ 
has been adopted. 

\begin{figure}
  \centering
  \includegraphics[width=\linewidth]{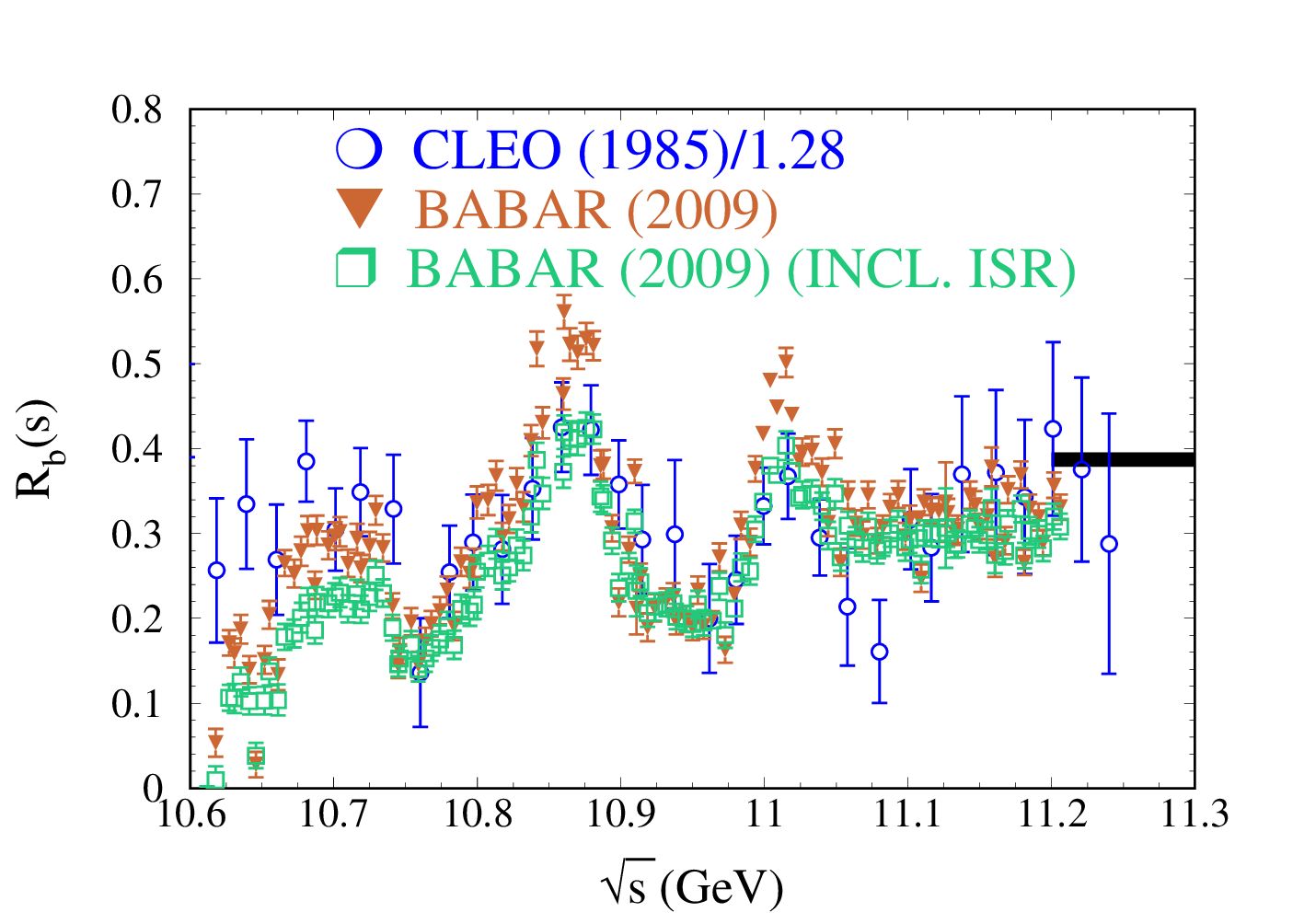}
  \caption[]{Comparison of rescaled CLEO data for $R_b$ with BABAR data
    before and after deconvolution. The black bar on the right corresponds to
    the theory prediction~\cite{Harlander:2002ur}.} 
  \label{fig:R_b}
\end{figure}

In Fig. \ref{fig:R_b} we show the BABAR  data \cite{:2008hx} (after
subtraction of the radiative tails of $\Upsilon(1S)$ to
$\Upsilon(4S)$), together with $R_b$ after deconvolution of ISR 
and correcting for the running of $\alpha(s)$. We also
show the CLEO data \cite{Besson:1984bd} after the aforementioned rescaling.

It is now straightforward to evaluate the contribution to the moments.
The result is listed in Tab.~\ref{tab:Mn5} and compared to our earlier analysis
based on the CLEO result~\cite{Besson:1984bd}. The error of this new result is  
dominated by the correlated systematic error of the BABAR measurement
which amounts to about 3.5\%. In addition we use a 2\% error for the uncertainty
from the details of the matching between the tail of $\Upsilon(4S)$ and
the continuum around $\sqrt{s} = 10.62~\text{GeV}$, which we add in quadrature. 

As is evident from Tab.~\ref{tab:Mn5}, the agreement between old and new
result  
is remarkable giving additional confidence in the procedure used in
Ref.~\cite{Kuhn:2007vp}.
The new experimental input and the new information on the
coefficients $\bar C^{30}_n$  lead to a significant
reduction of the error on $m_b$, as shown below.

\begin{table}[t]
\begin{center}
{
{\scalefont{0.9}
\begin{tabular}{l|llll}
$n$                                                 
& 1           & 2           & 3           & 4 \\\hline
${\cal M}_{n,\  {\rm old}}^{\rm dat}\times 10^{(2n+1)}$  
& $0.296(32)$ & $0.249(27)$ & $0.209(22)$ & $0.175(19)$\\ 
${\cal M}_{n,\  {\rm new}}^{\rm dat}\times 10^{(2n+1)}$  
& $0.287(12)$ & $0.240(10)$ & $0.200(8)$ & $0.168(7)$\\ 
\hline
${\cal M}_{n,\  {\rm new}}^{\rm exp}\times 10^{(2n+1)}$  
& $4.592(31)$ & $2.872(28)$ & $2.362(26)$ & $2.170(26)$\\ 
  \end{tabular}
}
}
\caption{\label{tab:Mn5}Moments in $(\text{GeV})^{-2n}$ for the bottom quark
  system from the 
  threshold region $\langle 10.62$ GeV, $11.24\ \text{GeV}\rangle$
  from Ref.~\cite{Kuhn:2007vp} (old) and this letter (new).
  Also the new total experimental moments are given.
}
\end{center}
\end{table}


\section{IV. Quark Masses}
\label{sec:mass}

In the absence of new data the analysis of $m_c$ will be based on the moments
listed in Tab.~6 of Ref.~\cite{Kuhn:2007vp}. As emphasized earlier \cite{Kuhn:2001dm,Kuhn:2007vp} it is convenient
to  consider as primary quantity the running quark mass at scale 3~GeV. 
This is the natural
scale for the sum rule (corresponding roughly to the charm
threshold) and, as a consequence of the smaller strong
coupling  constant, the perturbative series exhibits a more stable behaviour.

If not stated otherwise, all input parameters and assumptions are
identical to those of Ref.~\cite{Kuhn:2007vp}.
In particular we adopt $\alpha_s(M_Z)=0.1189$.
The new results and the corresponding errors are listed in
Tab.~\ref{tab:mc1}. Compared to \cite{Kuhn:2007vp}, the shift induced by the 
analytic results for $\bar{C}_n^{(30)}$ amounts to 3, 4 and 8~MeV for
n=2, 3 and 4 respectively.
The results for all four moments are nicely consistent, and the three lowest
moments exhibit comparable errors, ranging  between 13~MeV and 17~MeV. 
Note, that the
relative composition of the experimental input varies strongly from low to
high moments: For $n=1$ the contributions from narrow resonances and continuum 
are roughly comparable, for $n=3$ the continuum contribution amounts to 
about 10\%. 
Furthermore, the experimental contribution to the error decreases with
increasing $n$ from 9~MeV to 5~MeV, the $\mu$-dependence, reflecting the theory
uncertainly, increases from 2~MeV to 7~MeV. Despite the significant
differences in the composition of the errors, the results are perfectly
consistent. Since the result 
from $n=1$ has the smallest dependence on the strong coupling and the smallest
total error we take as our final value 
\begin{equation}
\label{eq:mc_3}
m_c(3~\text{GeV}) = 986(13)~\text{MeV}\,,
\end{equation}
and consider its  consistency with $n=2$, $3$ and $4$ as additional
confirmation.

\begin{table}[t]
\begin{center}
{
\begin{tabular}{l|r@{}l|rrrr|c}
\hline
$n$ & \multicolumn{2}{|c|}{$m_c(3~\text{GeV})$} & 
exp & $\alpha_s$ & $\mu$ & np & 
total
\\
\hline
       1&   &986&  9&   9&   2&  1 & 13 \\
       2&   &976&  6&  14&   5&  0 & 16 \\
       3&   &978&  5&  15&   7&  2 & 17 \\
       4&  1&004&  3&   9&  31&  7 & 33 \\
\hline
\end{tabular}
}
\caption{\label{tab:mc1}Results for $m_c(3~\text{GeV})$ in MeV
  obtained from Eq.~\eqref{eq:m_Q}. The errors are from experiment,
  $\alpha_s$, variation of $\mu$
  and the gluon condensate.
}
\end{center}
\end{table}

Transforming this to the scale-invariant mass $m_c(m_c)$~\cite{Chetyrkin:2000yt}, including the
four-loop coefficients  of the renormalization group functions one
finds~\footnote{In Ref.~\cite{Kuhn:2007vp} an imprecise routine has been
  used to compute the scale-invariant charm quark mass with the result
  $m_c(m_c)=1286$~MeV for the central value. In this letter we solve the
  coupled system of differential equations numerically which has also been
  done in Ref.~\cite{Allison:2008xk} where $m_c(m_c)=1268$~MeV has been
  obtained using $\alpha_s(M_Z)=0.1176$.}
$m_c(m_c)=1279(13)~\text{MeV}$. Let us recall at this point that a recent
lattice determination, 
combining a lattice simulation for the data for the pseudoscalar correlator
with the perturbative  three- and four-loop result
\cite{Chetyrkin:1997mb,Sturm:2008eb,Maier:2009} has led to
$m_c(3~\text{GeV})= 986(10)$~MeV \cite{Allison:2008xk} in remarkable
agreement with \cite{Kuhn:2007vp} and the present analysis.

The same approach is also applicable for the case of the bottom quark. 
Using the new moments with their significantly reduced experimental
error (see Tab.~\ref{tab:Mn5}), one obtains the results 
for the bottom quark mass at the scale $\mu=10\,$GeV as listed in
Tab.~\ref{tab::mb}. In comparison with the previous determination a minute 
upwards shift of 1~MeV (resulting from an upward shift of +3~MeV from
the new data and a downward shift of $-$2~MeV from the new theory input) 
and a reduction of both experimental and theory error is observed. 
The three results based on $n=1$, $2$ and $3$ are of comparable
precision. The relative size of the contribution from the continuum above 11.24
GeV which is modelled by perturbative QCD decreases for the higher moments $n=2$
and 3. On the other hand the theory uncertainty, exemplified by the $\mu$
dependence is still acceptable. We therefore adopt the result from $n=2$ (which
is roughly between the $n=1$ and $n=3$ values and which exhibits the smallest 
error) as our final result
\begin{align}
m_b(10~\text{GeV}) =&\,3610(16)~\text{MeV}\,, \notag\\
m_b(m_b) =&\,4163(16)~\text{MeV}\,.
\label{eq:mb}
\end{align}
These values are well consistent with the previous determination \cite{Kuhn:2007vp}
$m_b(10~\text{GeV})=3609(25)$~MeV and $m_b(m_b)= 4164(25)$~MeV.

\begin{table}[t]
\begin{center}
{

\begin{tabular}{l|l|rrr|l|l}
\hline
$n$ & $m_b(10~\text{GeV})$ & 
exp & $\alpha_s$ & $\mu$ &
total &$m_b(m_b)$
\\
\hline
        1&  3597&  14&  7&   2&  16&  4151 \\
        2&  3610&  10&  12&  3&  16&  4163 \\
        3&  3619&  8&  14&   6&  18&  4172 \\
        4&  3631&  6&  15&  20&  26&  4183 \\
\hline
\end{tabular}
}
\caption{\label{tab::mb}Results for $m_b(10~\text{GeV})$ and $m_b(m_b)$ 
in MeV
  obtained from Eq.~\eqref{eq:m_Q}.
  The errors are from experiment, $\alpha_s$ and the variation of $\mu$.
}
\end{center}
\end{table}

It is straightforward to evolve the new value for $m_b$ to the normalization
point at $M_Z$ and $m_t(m_t)= 161.8~\text{GeV}$
\begin{align}
  m_b(M_Z) =& 2835 \pm 13 \pm 17~\mbox{MeV} \,,\notag\\
  m_b(161.8~\mbox{GeV}) =& 2703 \pm  12 \pm 19~\mbox{MeV}
  \,,
\end{align}
where a matching to the $n_f=6$ flavour theory has been performed in order to
arrive at $m_b(161.8~\mbox{GeV})$.
The first error originates from Eq. \eqref{eq:mb} the second from
$\delta\alpha_s$.

For some of the applications it might be useful to explicitely exhibit the
$\alpha_s$ dependence of our result, which is given by
\begin{align}
  m_c(3~\text{GeV}) =& \left(986-\frac{\alpha_s-0.1189}{0.002}\cdot 9
  \pm 10\right)~\text{MeV}\,,\notag\\
  m_b (10~\text{GeV}) =& \left(3610  -\frac{\alpha_s-0.1189}{0.002}\cdot 12 
  \pm 11\right)~\text{MeV}\,,\notag\\
  m_b(m_b) =& \left(4163  +\frac{\alpha_s-0.1189}{0.002}\cdot   7
  \pm 14\right)~\text{MeV}\,,\notag\\
  m_b(M_Z)= &\left(2835 -\frac{\alpha_s-0.1189}{0.002}\cdot 27
  \pm 8\right)~\text{MeV}\,,\notag\\
  m_b(161.8~\mbox{GeV})=& \left(2703 -\frac{\alpha_s-0.1189}{0.002}\cdot 28
  \pm 8 \right)~\text{MeV}\,,\notag\\
\label{eq:as_dep}
\end{align}
where $\alpha_s=\alpha_s(M_Z)$.
When considering the ratio of charm and bottom quark masses,  part of
the  $\alpha_s$ and of the $\mu$ dependence cancels
\begin{equation}
  \label{eq:m_ratio}
  \frac{m_c(3~\text{GeV})}{m_b(10~\text{GeV})}=
  0.2732 -\frac{\alpha_s-0.1189}{0.002}\cdot 0.0014 \pm 0.0028 
  \,,
\end{equation}
which might be a useful input in ongoing analyses of bottom decays.

\section{V. Summary}
\label{sec:sum}

Based on new four-loop results for the higher derivatives 
of the vacuum polarization function and new BABAR data 
for bottom quark production in the threshold region, a 
reanalysis of the charm- and bottom-quark mass determination 
has been performed. The new data, a posteriori, give 
additional support to the analysis of CLEO data 
presented in Ref.~\cite{Besson:1984bd} and, furthermore, lead to a 
significant reduction of the experimental error. 
The new theory results for the higher moments lead to a 
further reduction of the theory uncertainty and, 
equally important, demonstrate the consistency between the    
analysis based on different moments. The final results,
$m_c(3 {\rm GeV})= 0.986(13)~{\rm GeV}$ and 
$m_b(10{\rm GeV}) =3.610(16)~{\rm GeV}$ are consistent with the earlier 
determination in Ref.~\cite{Kuhn:2007vp} and, together with
Ref.~\cite{Allison:2008xk}, constitute the most precise
determination of charm- and bottom-quark masses to date.

\section*{Acknowledgments}
\label{sec:ack}

This work was supported by the Deutsche Forschungsgemeinschaft through
the SFB/TR-9 ``Computational Particle Physics''. Ph.\,M. was supported
by the Graduiertenkolleg ``Hochenergiephysik und Teilchenastrophysik''.
A.\,M. thanks the Landesgraduiertenf\"orderung for support.
The work of C. S.
was supported by U.S. DOE under contract No. DE-AC02-98CH10886.

\bibliographystyle{hep}
\bibliography{hep}

\end{document}